\documentclass[aps,preprint,showpacs]{revtex4-1}

\usepackage{graphicx}
\usepackage{subcaption}
\usepackage{dcolumn}
\usepackage{bm}
\usepackage{float}
\usepackage{amsmath}

\begin{document}
	
	\title{Carrier Dynamics in a Tunneling Injection Quantum Dot 
		Semiconductor Optical Amplifier}
	\author{I. Khanonkin}
	\email{ikhanonkin@technion.ac.il}
	\affiliation{Andrew and Erna Viterbi Department of Electrical Engineering, Technion, Haifa 32000, Israel}
	\author{M. Lorke}
	\affiliation{Institute for Theoretical Physics, University of Bremen, 28334 Bremen, Germany}
	\author{S. Michael}
	\affiliation{Institute for Theoretical Physics, University of Bremen, 28334 Bremen, Germany}
	\author{A. K. Mishra}
	\affiliation{School of Physical and Mathematical Sciences, Nanyang Technological University, Singapore 637371}
	\author{J. P. Reithmaier}
	\affiliation{Institute of Nanostructure Technologies and Analytics, Technische Physik, CINSaT, University of Kassel, Kassel 34132, Germany}
	\author{F. Jahnke}
	\affiliation{Institute for Theoretical Physics, University of Bremen, 28334 Bremen, Germany}
	\author{G. Eisenstein}
	\affiliation{Andrew and Erna Viterbi Department of Electrical Engineering, Technion, Haifa 32000, Israel}
	
	\begin{abstract}
		The process of tunneling injection is known to improve the dynamical characteristics of quantum well and quantum dot lasers; in the latter, it also improves the temperature performance. The advantage of the tunneling injection process stems from the fact that it avoids hot carrier injection, which is a key performance-limiting factor in all semiconductor lasers. The tunneling injection process is not fully understood microscopically and therefore it is difficult to optimize those laser structures. We present here a numerical study of the broad band carrier dynamics in a tunneling injection quantum dot gain medium in the form of an optical amplifier operating at 1.55 $\mu$m. Charge carrier tunneling occurs in a hybrid state that joins the quantum dot first excited state and the confined quantum well - injection well states. The hybrid state, which is placed energetically roughly one LO phonon above the ground state and has a spectral extent of about $5~meV$, dominates the carrier injection to the ground state. We calculate the dynamical response of the inversion across the entire gain spectrum following a short pulse perturbation at various wavelengths and for two bias currents. At a high bias of $200~mA$, the entire spectrum exhibits gain; at $30~mA$, the system exhibits a mixed gain - absorption spectrum.  The carrier dynamics in the injection well is calculated simultaneously.  We discuss the role of the pulse excitation wavelengths relative to the gain spectrum peak and demonstrate that the injection well responds to all perturbation wavelengths, even those which are far from the region where the tunneling injection process dominates.
		
	\end{abstract}
	
	\pacs{}
	
	\maketitle
	
	\section{Introduction}
	
	The most basic mechanism limiting the modulation capabilities of semiconductor lasers is the gain nonlinearity, which originates from several processes including hot carrier injection \cite{Tucker1985,hall1994subpicosecond,mecozzi1997saturation}. An attractive way to diminish the hot carrier effect is to employ a delta doped film near the active region \cite{buchinsky1997n} or a tunnelling injection (TI) structure \cite{Bhattacharya1993}, which feeds cold carriers from an injection well (IW) reservoir directly to the lasing state. The successful use of TI was demonstrated for quantum well lasers more than twenty years ago \cite{Bhattacharya1993} and later for quantum dot (QD) lasers at short wavelengths \cite{ghosh2002dynamic} as well as at $1550~nm$ \cite{bhowmick2014high}. The TI concept can also improve the temperature stability \cite{asryan2001tunneling}.
	
	The advantages associated with the TI process were demonstrated often, however, the exact microscopic details of the tunneling process in a TI-QD laser are not fully understood and therefore it is hard to optimize QD laser structures that realize the full potential of the TI concepts. Different aspects of TI-QD lasers were modeled previously \cite{Bhattacharya1996,asryan2001tunneling,gready2011effects,gready2010carrier}. These models were generally based on a rate equation formalism where the tunneling process serves as an additional carrier injection path with a known rate which may depends on bias \cite{gready2011effects,gready2010carrier}. The tunneling process itself was also studied in terms of phonon assisted carrier transfer \cite{chang2004phonon,mielnik2015phonon}.
	
\begin{figure*}[htb]
		\centering
		\includegraphics[width=6cm]{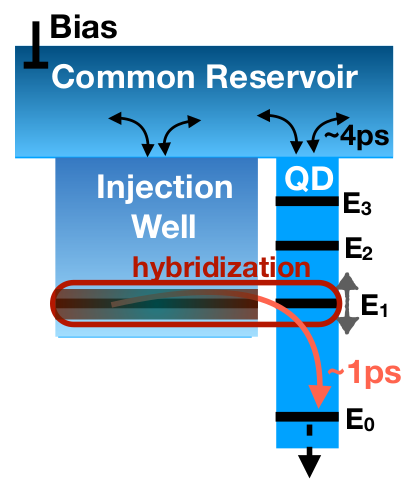}
		\caption{Energy band structure of a TI-QD SOA for a $4.5~nm$ wide IW and QDs with in plane dimensions of 16x24 nm and a height of 3 nm. The values of the energy levels are $-13.35, -41.15,-73.62~and~ -133.80~meV$ below the reservoir level. The tunneling process from the IW to the QD ground state is shown schematically. The conventional cascade carrier relaxation process from the common reservoir to the QD ground state takes $4-5~ps$. The tunneling injection design shortens the recovery process to $1-1.2~ps$} and diminishes all other carrier capture process to the ground state.
		\label{scetch} 
	\end{figure*}

	A recent paper by Michael et al., \cite{michael2018interplay} addressed the tunneling process in a TI-QD structure by introducing a joined (hybrid) energy state that couples the first excited state of the QD with the confined quantum well (QW) injection reservoir (IW) levels and has an energy extent of roughly $5~meV$. The energy levels of that structure were calculated using a  $k \cdot p$ model and the scattering rates were found by conventional overlap of the various wave functions. The energy band diagram and the most important scattering rates are shown in Fig. \ref{scetch}.
	
	A detailed study of the carrier dynamics is best done in a semiconductor optical amplifier (SOA) since it is a single pass device and therefore the responses are not masked by resonances from end facets. A single wavelength pump-probe  experimental analysis of a TI-QD SOA operating at 1300~nm was demonstrated in \cite{pulka2012ultrafast} and showed a fast recovery of up to 10~ps.
	
	We describe here a simulation of the broad band dynamic response of a TI-QD SOA following a perturbation by a short pulse. Using the model described in \cite{capua2013finite} with modifications based on \cite{khanonkin2017ultra}, we calculate the evolution of the inversion at every point along the SOA and across its entire gain spectrum. Additionally, we calculate the carrier dynamics in the IW. The calculations correspond to two bias levels and three pump pulse wavelengths. We find that the IW responds dynamically  even for pump wavelengths which are far from the spectral region where the hybrid state dominates. 
	
	\section{Simulation pump-probe model of the TI-QD SOA}
	The theoretical investigation of ultra-short pulses propagation in a QD amplifier is based on a semiclassical description of the light-matter interaction \cite{icsevgi1969propagation}, solved in the dipole moment approximation. We employ a numerical finite-difference time-domain model, developed in \cite{capua2013finite,mishra2015coherent,karni2015nonlinear}, that solves Lindblad equations for the occupation probabilities of a cascade of two-level quantum systems having different transition energies, that represent the inhomogeneously broadened ensemble of QDs. Simultaneously, it solves Maxwell’s equation for the electromagnetic field of the propagating pulse, where the vector polarization includes contributions from the interaction with the QDs, from two-photon absorption (TPA) and its accompanying Kerr-like effect as well as from group velocity dispersion (GVD) and the refractive index dependence on the carrier population, known as the plasma effect \cite{karni2015nonlinear}.
	
	We analyze an InAs/InP  TI QD SOA operating at 1.55 $\mu$m whose epitaxial structure and dimenssions are described in \cite{bauer2018growth}. A hybrid state that couples the first excited state of the QD and the bottom of the IW continuum of states  \cite{michael2018interplay}  ensures a fast carrier capture with a typical time constant of about 1 ps. Efficient formation of the hybrid state requires a $4-5~nm$ wide  quantum well IW, a narrow barrier between the QDs and the IW and a first QD excited state that is energetically located roughly one longitudinal phonon (LO) from its ground state. Details are presented in the appendix. 
	
\begin{figure*}[htb]
		\centering
		\includegraphics[width=8.6cm]{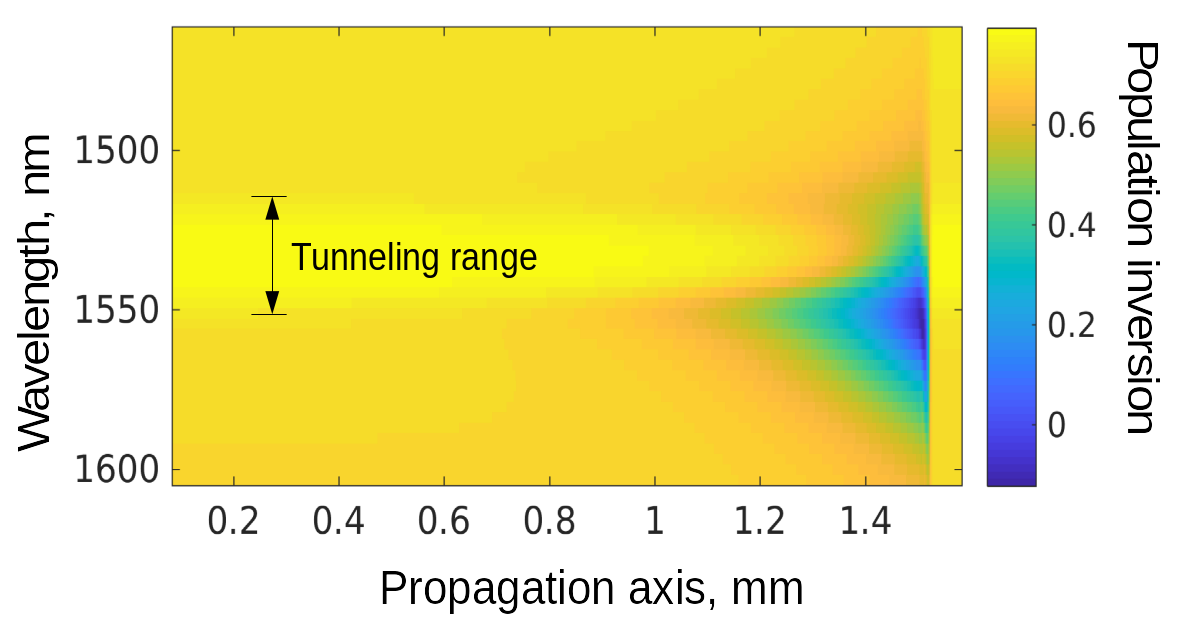}
		\caption{ Spatial evolution of the population inversions across the QD spectrum after a perturbation by a $100~pJ$ optical pulse at $1530~nm$. The amplifier is biased at $200~mA$ where it exhibits high gain across its entire gain spectrum.} 
		\label{3D} 
	\end{figure*}

	In addition to an accelerated replenishment rate of carriers in the QD ground state, the TI process adds an attractive feature by which it discriminates between groups of QDs and in fact filters the inhomogeneously broadened gain spectrum \cite{bhattacharya2003carrier}. This is due to the finite spectral extent of the hybrid state. QDs that overlap the hybrid state spectral range are fed mainly by carriers originating in the IW with a negligible contribution from the common high energy reservoir. In contrast,  QDs with transition energies far from the tunneling range, are fed mainly from the reservoir by a cascade process. This is shown in Fig. \ref{3D} which shows the population inversion (for a large bias, a $100~pJ$ pump pulse and at all wavelengths) behind a pulse that propagated up to the output facet. The occupation probabilities of QDs that are spectrally close to the gain peak ($1530~nm$) where the tunneling process is most efficient, is high. The effective capture rate to the ground state is determined by a wavelength dependent tunneling rate which is modeled by a Gaussian profile with a variance of $5~meV$. Details are described in the appendix. Efficient tunneling makes the relaxation processes of direct and cascaded capture and escape, from and to the common reservoir, insignificant. However, for wavelengths far from the tunneling range, these conventional processes dominate. The two regimes are combined in the model by including a wavelength dependent direct relaxation rate which has a Gaussian profile with an opposite sign that of the tunneling time constant. Relaxation from the reservoir has a very long time constant, $21~ps$ in the tunneling spectral range which shortens to $4~ps$ at the gain spectrum edges.
		
	The inversion profile shown in Fig. \ref{3D} is for an amplifier biased at $200~mA$ and a $100~pJ$ pulse which is injected at $1530~nm$. The pulse red-shifts upon propagation due to various nonlinear effects \cite{karni2015nonlinear}. The figure shows clearly that in QDs which are close to the gain peak, the inversion recovers faster than in QDs which are far from the peak (on the long wavelength side). This is a clear indication of the faster carrier replenishment in the spectral range where the TI process is efficient.
	
\begin{figure}
		\centering
		\begin{subfigure}[b]{0.48\textwidth}
			\includegraphics[width=8cm]{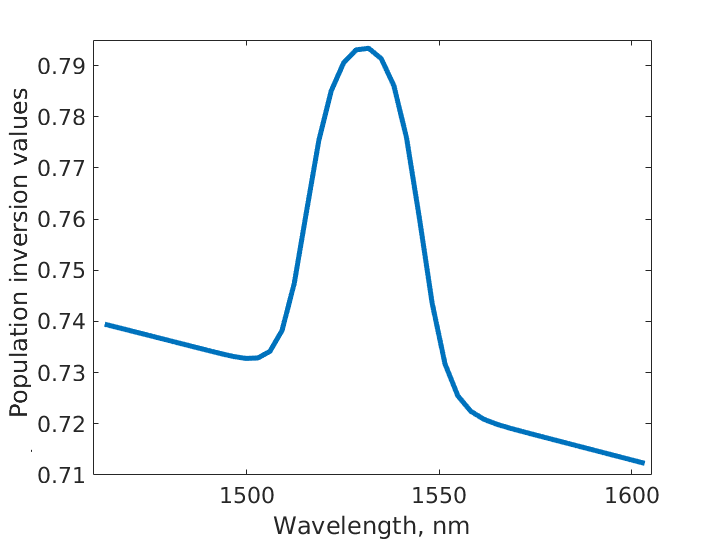}
			\caption{}
			\label{Steady_State_30mA_2D} 
		\end{subfigure}
		~ 
		\begin{subfigure}[b]{0.44\textwidth}
			\includegraphics[width=8.5cm]{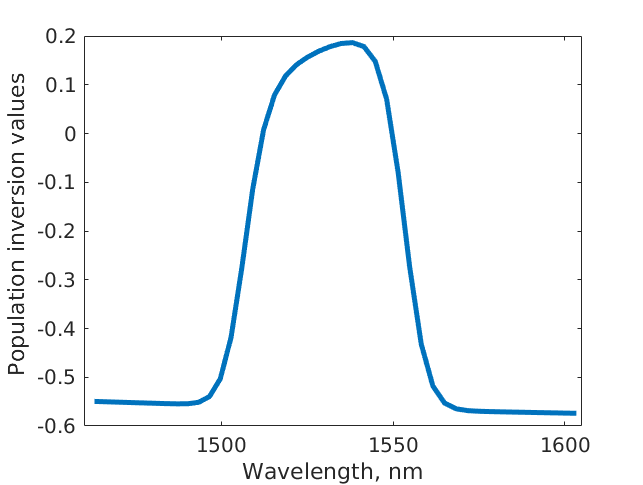}
			\caption{}
			\label{Steady_State_200mA_2D} 
		\end{subfigure}
		\caption{Steady state inversion spectra of the QD population for a bias of (a)~200~mA~(b)~30~mA.}
		\label{Steady_states}
	\end{figure}

	We investigate carrier dynamics in the TI QD SOA for two bias levels: $200~mA$, where the amplifier exhibits high gain at all wavelengths and $30~mA$, where the QDs overlapping the tunneling range have gain while QD far from the peak are absorbing. Wavelength dependent inversion in the un-perturbed steady states, are shown in Fig. \ref{Steady_states}  for the two bias levels. We probe the SOA response  for input pulses near the gain peak ($1530~nm$) and out of the tunneling range, $1480~nm$ and $1590~nm$.  Using the refractive index information, the spatial distribution of the charge carriers was translated to a time evolution, which reflects the commonly measured time domain response of the probe transmission \cite{khanonkin2017ultra}. 
	
	\section{Carrier dynamics in a TI-QD SOA biased to the gain regime}
	
\begin{figure*}[htb]
		\centering
		\includegraphics[width=8.6cm]{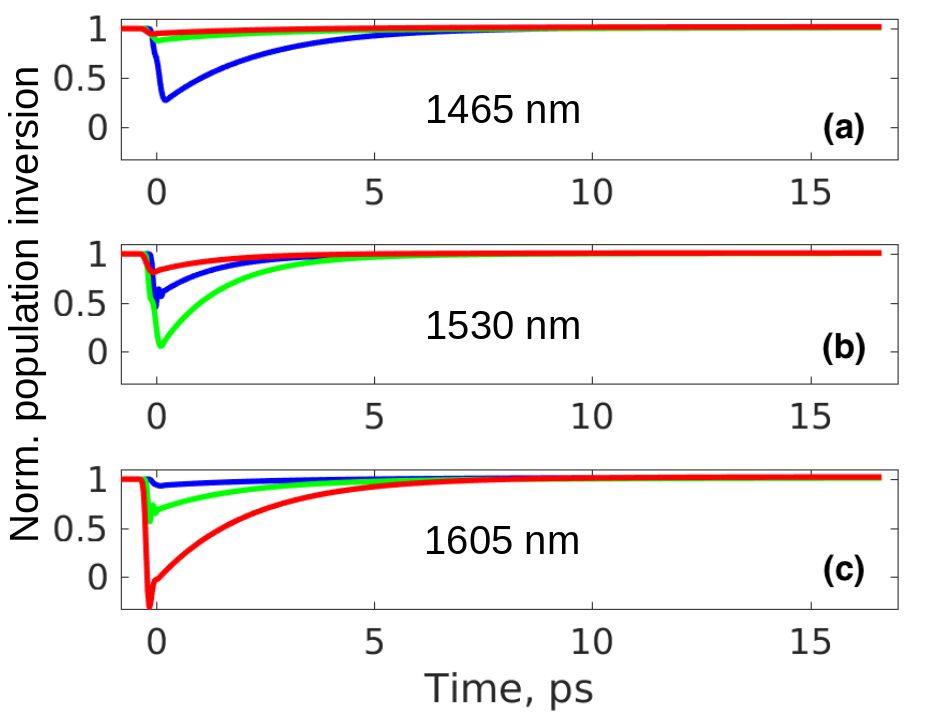}
		\caption{Time evolution of the inversion following the perturbation at different pump wavelengths: $1480$ (blue trace), $1530$(green trace) and $1590~nm$ (red trace). The QD ground state inversions are normalized to their respective values prior to the perturbation. The inversions are probed at (a) - 1465 nm, (b) - 1530 nm, (c) -1590 nm. The SOA  is biased at 200 mA where it exhibits high gain across its entire gain spectrum.}
		\label{Time_dependence_200mA} 
	\end{figure*}

	Time dependent responses of the QD ground state inversions probed at different spectral location are shown in the Fig. \ref{Time_dependence_200mA} for a bias of 200 mA and a pump energy of $200~pJ$. Each curve represents the response of a particular group of QDs. Colored curves indicate different input pump wavelengths (blue - $1480~nm$, green - $1530~nm$, red -$1590~nm$). The traces are normalized to their respected values prior to the perturbation. The initial depth in the inversion at different pump wavelength is directly related to the induced depletion by the corresponding pump. 

Fast recovery in the QD that spectrally overlap the tunneling region, is clearly seen in Fig. \ref{Time_dependence_200mA} (b). The initial saturation is deepest and the recovery is fastest for a pump wavelength of $1530~nm$ but even for pump wavelengths that are far from the TI regime, the QDs near the gain peak have a significant dynamical response since even the slightest depletion of the ground state or the reservoir is easily sensed and corrected for by the efficient tunneling process.

The situation is different for probes at the spectral edges. At $1465~nm$, Fig. \ref{Time_dependence_200mA} (a), the inversion responds only for a pump at the same wavelength and is indifferent to pumps at lower energies. In contrast, the QDs at $1605~nm$, Fig. \ref{Time_dependence_200mA} (a), exhibit some dynamical response for the higher energy pump pulses. Those pump pulses red shift upon propagation and approach the tail of the long wavelength probe region and hence affect it.
	
\begin{figure*}[htb]
		\centering
		\includegraphics[width=14cm]{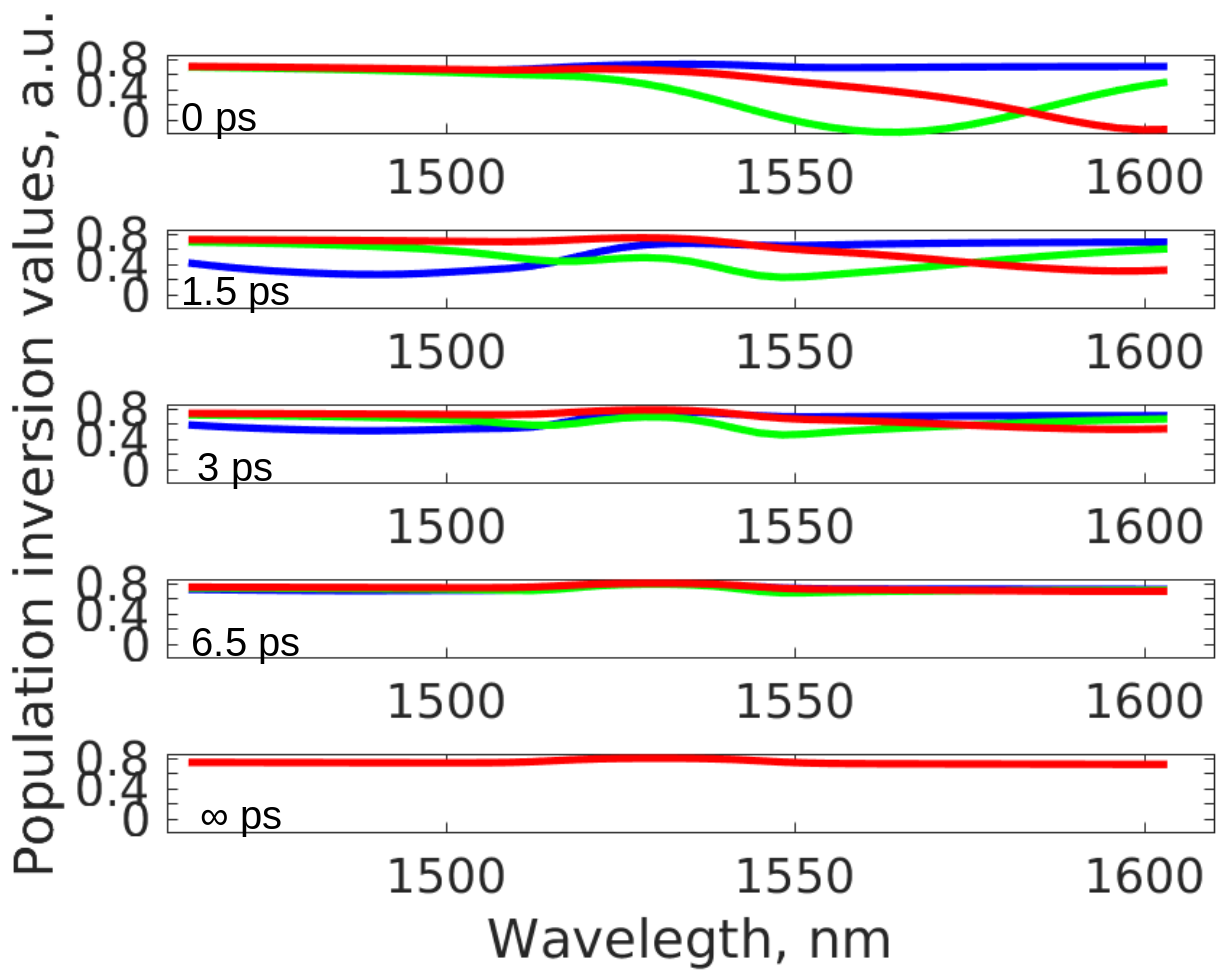}
		\caption{Time resolved spectral profiles of the ground state recovery following the perturbation at different pulse wavelengths: $1480$ (blue trace), $1530$(green trace) and $1590~nm$ (red trace). The SOA  is biased at 200 mA where it exhibits high gain across its entire gain spectrum.}
		\label{Spectrum_dependence_200mA} 
	\end{figure*}

	Fig. \ref{Spectrum_dependence_200mA} describes time resolved spectra of the QD ground state inversion. The dip in the inversion right after the interaction with the pump is slightly shifted towards long-wavelength QDs since the pump red-shifts during propagation. The 1480 nm trace appears at $0~ps$ with no perturbation since the pump is chirped during the propagation and has not yet reached the short-wavelength QDs.  Since the QDs close to 1530 nm are replenished faster, the inversion curves at $1.5~ps$ and $3~ps$ show a wavy profile. At very long times, all QDs recover to their steady state profiles.  
	
\begin{figure*}[htb]
		\centering
		\includegraphics[width=8.6cm]{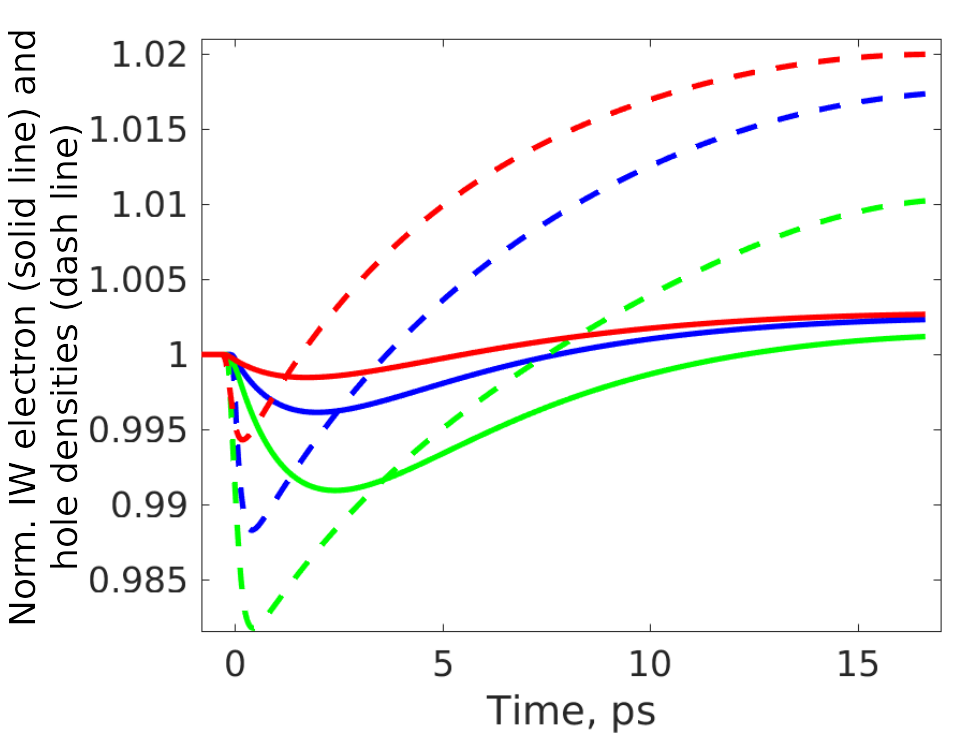}
		\caption{Injection well electron (solid line) and hole (dashed line) densities for the three perturbations of different spectral locations. The SOA is biased at 200 mA where it exhibits high gain across its entire gain spectrum.}
		\label{FIG_IW_e_h_200mA} 
	\end{figure*}

	Fig. \ref{FIG_IW_e_h_200mA} shows the response of the electron and hole densities in the IW for the same three perturbation wavelengths. Ground state carriers recover significantly faster when the perturbation is at $1530~nm$ which is within the TI spectral range. The perturbation at short-wavelength has a larger effect on the IW carrier density compared to the long wavelength since the red-shift experience by the pump upon propagation leads to some  overlap with the tunneling region.
	
	\section{Carrier dynamics in TI-QD SOA electrically biased to the gain-absorption state}
	
\begin{figure*}[htb]
		\centering
		\includegraphics[width=8.6cm]{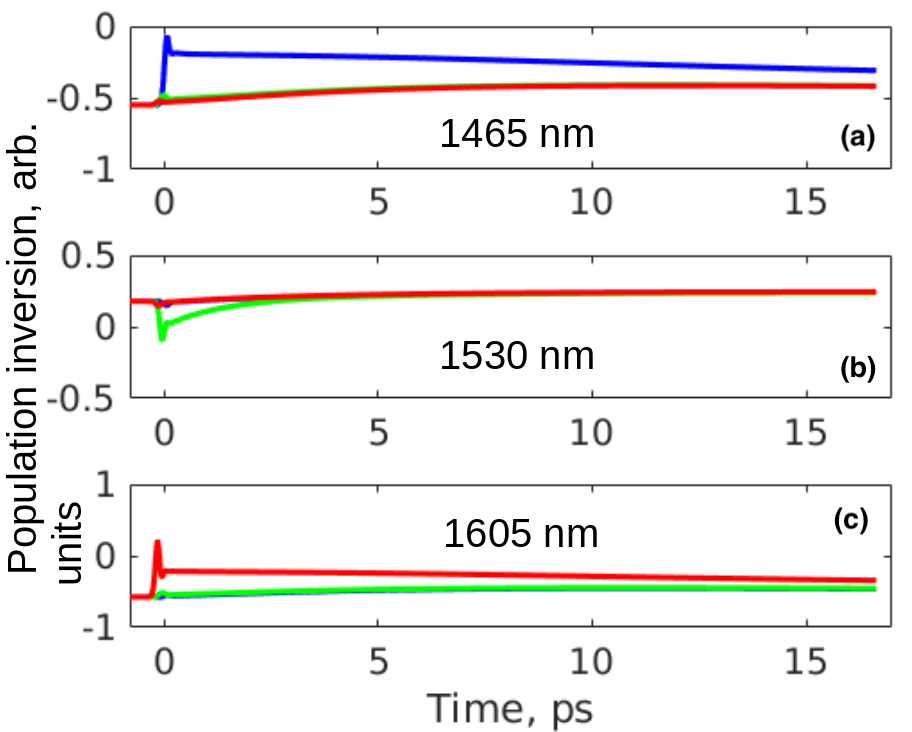}
		\caption{Time evolution of the inversion following the perturbation at different pump wavelengths: $1480$ (blue trace), $1530$(green trace) and $1590~nm$ (red trace). The QD ground state inversions are normalized to their respective values prior to the perturbation. The inversions are probed at (a) - 1465 nm, (b) - 1530 nm, (c) -1590 nm. The SOA is biased at 30~mA, where it exhibits a mixed gain-absorption state.}
		\label{FIG_Time_dependence_30mA} 
	\end{figure*}

	At low bias level, the  TI QD SOA experiences a mixed gain-absorption state. Specifically, close to 1530 nm, the QDs exhibit gain while towards the gain spectrum edges, the QDs are absorbing. The  un-normalized recovery traces in Fig. \ref{FIG_Time_dependence_30mA} show temporal responses to the three different pump wavelengths. The pump inverts the QDs located at the spectral edges when it is, correspondingly at the short and long wavelength side of the spectrum ( Fig. \ref{FIG_Time_dependence_30mA} (a) and (c), respectively). The responses of QDs at wavelengths far from the gain peak are all but diminished for other pump wavelengths while the QDs that participate in the tunneling process are replenished fast (green trace in Fig. \ref{FIG_Time_dependence_30mA} (b)). An increase of the occupation probabilities to above the level prior to the pulse arrival is a result of TPA that occur on a time scale of a few picoseconds .	
	
\begin{figure*}[htb]
		\centering
		\includegraphics[width=14cm]{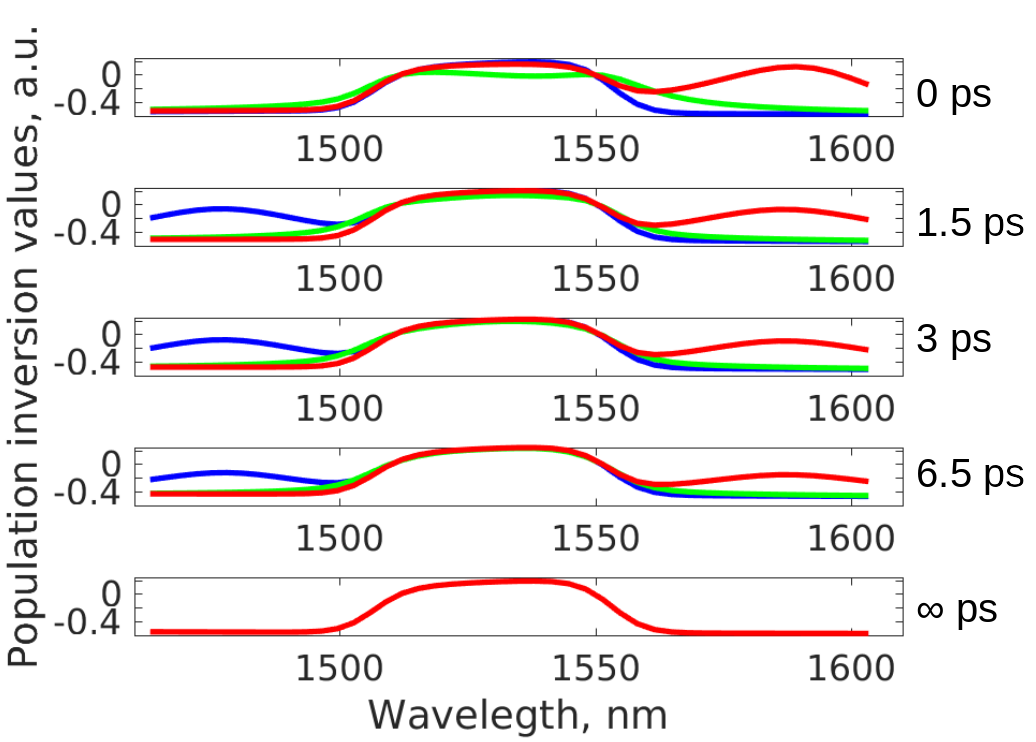}
		\caption{Time resolved spectral profiles of the ground state recovery following the perturbation at different pulse wavelengths: $1480$ (blue trace), $1530$(green trace) and $1590~nm$ (red trace). The SOA is biased at 30 mA, where it exhibits a mixed gain-absorption state.}
		\label{FIG_Spectrum_dependence_30mA} 
	\end{figure*}

	Time resolved spectral responses of the recovery are shown in Fig. \ref{FIG_Spectrum_dependence_30mA}. The  QDs at the gain peak experience a small perturbation for a pump centered at 1530 nm and a fast recovery to the steady state. In contrast, the time evolution of the recovery traces in the periphery of the QDs spectrum shows prolongated tails  that reach an equilibrium state on a rather long time scale.
	
\begin{figure*}[htb]
		\centering
		\includegraphics[width=8.6cm]{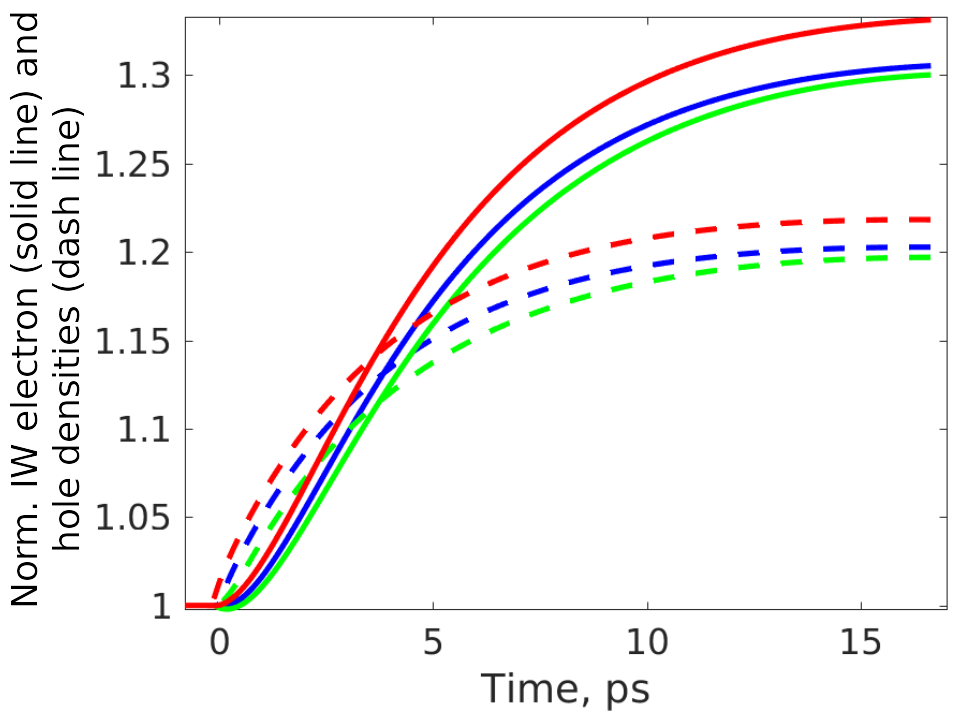}
		\caption{Injection well electron (solid line) and hole (dashed line) densities for the three perturbations of different spectral locations. The SOA  is biased at 30 mA where it exhibits a mixed gain-absorption state.}
		\label{FIG_IW_e_h_30mA} 
	\end{figure*}

	The IW electron and hole densities for the 30 mA bias level, Fig. \ref{FIG_IW_e_h_30mA}, are affected mainly by TPA.  The relative contribution of the TPA carriers that relax to the ground state is high compared to the equilibrium state when the bias level is low.

	\section{Conclusion}
	We investigated numerically carrier dynamics in TI QD SOA across its gain spectrum by simulating a broad band pump probe experiment. Fast carrier replenishment in the QDs within the spectral range of the tunneling injection process, is demonstrated. Moreover, we show various possibilities of amplifier operation; the amplifier can be biased in gain or in a mixed gain - absorption state. For those two different cases of amplifier operation, the effect of several perturbation wavelengths across the spectrum was analyzed. 
	
	\section{Appendix}
	The appendix describes a theoretical model of  an ultrashort-pulse propagation in the tunneling injection quantum dot amplifier. We simulate time and wavelength-resolved charge carrier dynamics in the inhomogeneously broadened ensemble of QDs by a set of rate equations. The QDs population is governed by relaxation from the common high energy carrier reservoir and from tunneling that originates in the injection well. The former occur via a cascaded transition through the excited states with a time constant of around 4 ps for InAs/InP QDs having dimensions of 16x24X3 nm. The reservoir is fed by an electrical DC source and by carriers generated through TPA that relax from higher energy levels. 
	
	\begin{gather*}
	\frac{\partial N_{IW}}{\partial t} = -\frac{ N_{IW}}{ \tau_{IW}} - \frac{ N_{IW_{targ}} \left(1- \exp \left(-\frac{N_{IW}}{N_{IW_{targ}}} \right) \right)}{ \tau_{tun} \cdot N_{D_{total}}} \frac{V_{QD}}{V_{IW}} \sum_{i=1}^{M} N_D^i \left(1-\rho_{11}^i\right) 
	\\
	+\frac{V_{QD}}{V_{IW}} \left(1- \frac{N_{IW}}{D_{IW}}\right) \sum_{i=1}^{M} 2N_{D}^{i} \frac{\rho_{11}^{i}}{\tau^i_{tun_{back}}}+\frac{V_{res}}{V_{IW}} \frac{N_{res}}{\tau_{cap_{IW}}} \left(1- \frac{N_{IW}}{D_{IW}}\right) - \frac{N_{IW}}{\tau_{esc_{IW}}}\left(1- \frac{N_{res}}{D_{res}}\right);
	\end{gather*}
	
	\begin{gather*}
	\frac{\partial P_{IW}}{\partial t}=-\frac{N_{IW}}{\tau_{IW}} +\frac{V_{res}}{V_{IW}} \frac{P_{res}}{\tau^h_{cap_{IW}}}  \left(1-\frac{P_{IW}}{D_{IW}} \right) - \frac{P_{IW}}{\tau^h_{esc_{IW}}} \left(1-\frac{P_{res}}{D_{res}}\right);
	\end{gather*}
	
	\begin{gather*}
	\frac{\partial N_{res}}{\partial t} = \frac{\eta_i J}{qd} - \frac{N_{res}}{\tau_{res}}+\frac{V_{TPA}}{V_{res}} \left( 1- \frac{N_{res}}{D_{res}}\right) \frac{N_{TPA}}{\tau_{TPA_{relax}}}
	\\
	- \frac{N_{res}}{N_{D_{total}} \cdot \tau_{d_{cap}}} \frac{V_{QD}}{V_{res}} \sum_{i=1}^{M} N_D^i \left( 1- \rho^i_{11}\right)
	+\frac{V_{QD}}{V_{res}} \left( 1- \frac{N_{res}}{D_{res}}\right) \sum_{i=1}^{M} 2N_D^i \frac{\rho_{11}^i}{\tau^i_{d_{esc}}} 
	\\
	- \frac{N_{res}}{\tau_{cap_{IW}}} \left( 1- \frac{N_{IW}}{D_{IW}}\right)+\frac{V_{IW}}{V_{res}} \frac{N_{IW}}{\tau_{esc_{IW}}} \left( 1- \frac{N_{res}}{D_{res}}\right);
	\end{gather*}
	
	\begin{gather*}
	\frac{P_{res}}{\partial t} = \frac {\eta_i J}{q d} - \frac{N_{res}}{\tau_{res}}  
	+\frac{V_{TPA}}{V_{res}} \left( 1- \frac{P_{res}}{D_{res}}\right) \frac{h_{TPA}}{\tau_{TPA_{relax}}} - \frac{P_{res}}{\tau_{cap}^h} \frac{V_QD}{V_{res}} \frac{1}{N_{D_{total}}} \sum_{i=1}^{M} N_D^i \rho_{22}^i
	\\ + \frac{V_QD}{V_{res}} \left( 1- \frac{P_{res}}{D_{res}}\right) \frac{1}{\tau_{esc}^h} \sum_{i=1}^{M} 2N_D^i \left( 1- \rho_{22}^i\right) - \frac{P_{res}}{\tau_{cap_{IW}}^h} \left( 1- \frac{P_{IW}}{D_{IW}}\right)+ \frac{V_{IW}}{V_{res}}\frac{P_{IW}}{\tau^h_{esc_{IW}}} \left( 1- \frac{P_{res}}{D_{res}}\right);
	\end{gather*}
	
	\begin{gather*}
	\frac{\rho^i_{11}}{\partial t}=- \gamma_c \rho_{11}^i +\frac{N_{res}}{2N_{D_{total}} \tau_{d_{cap}}} \left( 1- \rho_{11}^i\right) - \frac{\rho_{11}^i}{\tau^i_{d_{esc}}} \left( 1- \frac{N_{res}}{D_{res}} \right) -j \frac{\vec \mu \cdot \vec E}{\hbar} \left(\rho^i_{12} -\rho^i_{21}\right) \\+ \frac{N_{IW_{targ}} \left(1- exp \left(-\frac{N_{IW}}{N_{IW_{targ}}}\right)\right)}{\tau_{tun} \cdot 2 N_{D_{total}}} \left(1-\rho^i_{11}\right)-\left(1-\frac{N_{IW}}{D_{IW}}\right) \frac{\rho^i_{11}}{\tau^i_{tun_{back}}};
	\end{gather*}
	
	\begin{gather*}
	\frac{\partial \rho^i_{22}}{\partial t} = \gamma_c \rho^i_{11} - \frac{P_{res}}{\tau^h_{cap}} \frac{\rho_{22}^i}{2N_{D_{total}}}+\frac{\left( 1- \rho^i_{22}\right)}{\tau^h_{esc}} \left(1- \frac{P_{res}}{D_{res}}\right) + j \frac{\vec{\mu} \cdot \vec{E}}{\hbar} \left(\rho^i_{12}-\rho^i_{21}\right);
	\end{gather*}
	
	\begin{gather*}
	\frac{\partial \rho^i_{12}}{\partial t} = - \left( i \omega + \gamma_h \right) \rho^i_{12} - j \frac{\vec{\mu} \cdot \vec{E}}{\hbar} \left(\rho^i_{11}-\rho^i_{22}\right);
	\end{gather*}
	
	\begin{gather*}
	\frac{\partial N_{TPA}}{\partial t} = I_{TPA} - \frac{N_{TPA}}{\tau_{TPA_{relax}}} \left(1- \frac{N_{res}}{D_{res}}\right) - \frac{N_{TPA}}{\tau_{TPA_{rec}}};
	\end{gather*}
	
	\begin{gather*}
	\frac{\partial P_{TPA}}{\partial t} =I_{TPA} - \frac{P_{TPA}}{\tau^h_{TPA_{relax}}} \left(1- \frac{P_{res}}{D_{res}}\right) - \frac{N_{TPA}}{\tau^h_{TPA_{rec}}};
	\end{gather*} 
	where $N_{res}, P_{res}, N_{IW}, P_{IW}$ are, respectively, electron and hole densities in the corresponding reservoirs and IW. $J$ is the applied current density, $q$ is the electron charge. $\sum_{i=1}^{M}$ -represent summation over all energetically similar QDs and $N_D^i$ are their corresponding densities. $\rho^i_{11}~and~\rho^i_{22}$ are the occupation probabilities of the upper and lower states and $\rho^i_{12}=\left( \rho^i_{21} \right)^*$ are the coherence terms. $I_{TPA}, N_{TPA}~and~h_{TPA}$ are TPA carrier generation rate, electron and hole population in the TPA levels, respectively. The definition of other variables and their corresponding values are shown in table \ref{tabel}.
	
	\begin{table}[]
		\centering
		\caption{List of parameters}
		\label{tabel}
		\begin{tabular}{lll}
			$\tau_{cap_{IW}}$ & $1.5~ps$ & 3D to 2D electron capture time\\
			$\tau_{IW}$&$0.4~ns$ & IW lifetime \\
			$\tau^h_{cap_{IW}}=\tau^h_{esc_{IW}}$&$0.35~ps$ &  3D to 2D hole capture and escape times \\ 
			$\tau_{res}$&$0.4~ns$ & Reservoir lifetime \\
			$\tau_{TPA_{relax}}=\tau^h_{TPA_{relax}}$&$4~ps$ & Relaxation time of TPA carriers to the reservoir states \\
			$\tau_{TPA_{rec}}$ & 0.4 ns & TPA recombination time\\
			$\tau^h_{cap}=\tau^h_{esc}$ & $0.1~ps$ & 3D to 0D hole capture and escape times \\
			$\gamma_c$ & $2.5 \cdot 10^9 ~s^{-1}$ & Density matrix diagonal elements decay rate\\
			$\gamma_h$ &$2.85 \cdot 10^{12} ~s^{-1}$ & Density matrix off-diagonal elements decay rate\\
			$D_{IW}$& $1 \cdot 10^{25} ~ m^{-3}$ & 2D density of states \\
			$D_{res}$& $2 \cdot 10^{25} ~ m^{-3}$ & 3D density of states \\
			$\mu$& $0.5 \cdot 10^{-28} ~ C m$ & Dipole moment \\
			$\frac{V_{res}}{V_{QD}}$ & $3.5$ & 3D to 0D volume ratio\\
			$\frac{V_{IW}}{V_{QD}}$ & $1.15$ & 2D to 0D volume ratio\\
			$\frac{V_{TPA}}{V_{res}}$ & $50$ & TPA to reservoir volume ratio\\
			$N_{D_{total}}$& $4\cdot 10^{23}~m^{-3}$ & QD density of states\\
			$\eta_i$ & $0.5$ & Injection efficiency \\
			$L$ & $1.5~mm$ & SOA length\\
			$d$ & $18~nm$ & Active layer width comprised of 6 QD layers\\
		\end{tabular}
	\end{table}
	
	The tunneling time constant, $~\tau_{tun}$, is modeled  as Gaussian profile with a 5 meV variance. Its value at the gain peak is around 1 ps. Similar to the tunneling time constant, the wavelength-dependent capture time to the QD ground state $~\tau_{d_{cap}}$, is also assumed to have a Gaussian profile with a value of 4 ps outside the spectral range of tunneling process.
	
	The characteristic time constants of the carrier tunneling from QDs to IW,$~\tau_{tun_{back}}$, escape from QD to the common reservoir,$~\tau_{d_{esc}}$, and carrier escape from the IW states to reservoir, $~\tau_{esc_{IW}}$, are calculated according to the principle of detailed balance.
	
	We limit the maximum number of carriers, $N_{IW_{targ}}$, in the quantum well of IW, that participate in the tunneling process, to  $4.5\cdot 10^{23}~ m^{-3}$ \cite{michael2018interplay} since only the energetic bottom of the quantum well is hybridized with QDs.
	
	The Maxwell curl equations are solved simultaneously for the electromagnetic wave propagation is:
	\begin{gather*}
	\frac{\partial E_x}{\partial z} = -\mu_0 \frac{\partial H_y}{\partial t}, ~ -\frac{\partial H_y}{\partial z}=\frac{\partial D_x}{\partial t};
	\end{gather*}
	where $E_x, D_x, H_y$ are the electric field, electric displacement and magnetic field components, respectively, and $\mu_0$ is the permeability of the vacuum. Interaction with the material perturbs the polarization which includes several components:  radiation of the two-level systems, the plasma effect, dispersion, two photon absorption and its accompanying Kerr-like effect,. More details are given in \cite{karni2015nonlinear}.

	\section*{Funding Information}
	This work is partially supported by the Israel Science Foundation, grant number 1504/16.
	
	ML, SM and FJ acknowledge funding from the DFG and a grant for CPU time from the HLRN (Hannover/Berlin)

	\section*{Acknowledgments}

	\bibliography{sample}
	
\end{document}